\documentclass[superscriptaddress, twocolumn]{revtex4}
\usepackage{amsmath}
\usepackage{graphicx}

\usepackage{floatflt,epsfig}
\usepackage{rotating}
\begin{document}

\def\ii{\'{\i}}
\def\d{\mbox{d}}

\title{Slowly rotating pulsars}

\author{M.D. Alloy}
\affiliation{ Depto de F\ii sica - CFM - Universidade Federal de Santa
Catarina  Florian\'opolis - SC - CP. 476 - CEP 88.040 - 900 - Brazil}
\author{D.P. Menezes}
\affiliation{ Depto de F\ii sica - CFM - Universidade Federal de Santa
Catarina  Florian\'opolis - SC - CP. 476 - CEP 88.040 - 900 - Brazil}

\begin{abstract}
In the present work we investigate one possible variation on the usual
static pulsars: the inclusion of rotation. We use a formalism proposed 
by Hartle and Thorne to calculate the properties of rotating pulsars. 
All calculations were performed for zero temperature and fixed entropy 
equations of state. 
\end{abstract}

\maketitle

\vspace{0.5cm}
PACS number(s):26.60.+c,24.10.Jv,21.65.+f,95.30.Tg 
\vspace{0.5cm}

\section{Introduction}

Pulsars are believed to be the remnants of supernova explosions. They have 
masses 1--$2M_\odot$, radii $\sim10\rm\,km$, and a temperature of the 
order of $10^{11}\rm\,K$ at birth, cooling within a few days to about 
$10^{10}\rm\,K$ by emitting neutrinos. Pulsars are normally known as neutron 
stars. Qualitatively, a neutron star is analogous to a white dwarf star, 
with the pressure due to degenerate neutrons rather than degenerate 
electrons. The assumption that the 
neutrons in a neutron star can be treated as an ideal gas is not well 
justified: the effect of the strong force needs to be taken into 
account by replacing the equation of state (EoS) for an ideal gas by 
a more realistic EoS. The composition of pulsars  
remains a source of speculation, with some of the possibilities being 
the presence of hyperons \cite{aquino,magno,rafael}, a mixed phase 
of hyperons and quarks \cite{mp1,mp,pmp1,pmp2,trapping}, a phase of 
deconfined quarks or pion and kaon condensates \cite{kaons}. Another
possibility would be that pulsars are, in fact, quark stars \cite{quark}.
In conventional models, hadrons are assumed to be the true ground state of the 
strong interaction. However, it  has been argued 
\cite{itoh,bodmer,witten,haensel,olinto} that {\it strange matter}
composed of deconfined $u,d$ and $s$ quarks is the true ground  state of all 
matter. {\bf In the present work we have opted to use the term quark stars
  instead of strange stars because in more refined nuclear structure models, 
as the Nambu-Jona-Lasinio model for instance, the $s$ quark appears in a
  much smaller quantity than in the very naive MIT model, where it is
  considered in equal footing with $u$ and $d$ quarks \cite{quark} .}
In the stellar modeling, the structure of the star depends on the 
assumed EoS, which is different in each of the above mentioned cases. 
An important distinction 
between quark stars and conventional neutron stars is that the quark 
stars are self-bound by the strong interaction, whereas neutron stars 
are bound by gravity. This allows a quark star to rotate faster than 
would be possible for a neutron star \cite{quark, olinto}.

Once an adequate EoS is chosen, it is used as input to the 
Tolman-Oppenheimer-Volkoff (TOV) equations \cite{tov}, which are derived from 
Einstein's equations in the Schwarzschild metric for a static, 
spherical star. Some of the stellar properties, as the radius, gravitational
and baryonic masses, central energy densities, etc are obtained. These
results are then tested against some of the constraints provided by
astronomers and astrophysicists \cite{cottam,sanwal} and some of the EoS are
shown to be inappropriate for describing pulsars \cite{mp,pmp1,kaons}.

{\bf On the other hand, it is well known from the
Doppler broadening of the pulsars spectral lines that they rotate.
Some pulsars are observed to rotate with periods as small as
1.56 ms \cite{chakrabarty}. Newly born hot pulsars can rotate fast and hence
undergo instabilities.} The
effect of a rotating star on spacetime is commonly known as the Lense-Thirring
effect, which refers to the dragging of local inertial frames. Rotation breaks
spherical symmetry, but a rotating star is still axially symmetric. In this
case, the TOV equations are no longer valid. Hartle and Thorne proposed a 
perturbative approximation to treat rotating stars \cite{ht97}. The method was 
further developed and it is valid not only for slowly rotating stars, but also 
for stars rotating up to the Kepler frequency \cite{improved}. 
{\bf In a more recent work \cite{berti} the Hartle-Thorne approximation was 
tested against a full general relativistic numerical model available as part
of the numerical relativity library LORENE and it was shown that it is
reliable for most astrophysical applications. In \cite{berti} some EoS were
used to compute rotating star properties, all of them obtained at zero 
temperature.}

In the present work we use the Hartle-Thorne formalism to calculate the 
maximum mass, moment of inertia and eccentricity of all different classes of 
possible pulsars described above. In previous works \cite{antigo1,antigo2,
berti}, many EoS
have been investigated, but they were restricted to hadronic matter at
$T=0$. In the present paper we investigate all possible classes of pulsars 
(hadronic, hybrid and
quark stars) at zero and finite temperature, to account also for protoneutron
stars. It is important to distinguish between the EoS during the short time 
period when neutrinos are still trapped in the star, and the EoS after the 
neutrinos escape. Next we restrict ourselves to the second case, when pulsars
are believed to be already stable stars.
Notice also that the temperature in the interior of the star is 
not constant \cite{burrows,trapping}, but the entropy per baryon is. This is 
the reason for choosing fixed entropies to take the temperature effects into 
account. The maximum entropy  per baryon ($S$) reached in the core of 
a new born star is about 2 (in units of Boltzmann's constant) \cite{prak97}.
We then use EoS obtained with $S=0~(T=0$), 1 and 2.

Other important motivations for revisiting rotating stars is the fact that
as they slow down, the decreasing centrifugal force leads to increasing core
pressure and density. A softening of the EoS takes place and it can be the
result of a phase transition to quark matter \cite{pei}. If a first order
phase transition takes place at central densities, it was shown that the
moments of inertia behave in a characteristic way and the braking index
diverges \cite{heisel}. Gamma ray bursts are also linked with stellar phase 
transitions \cite{bursts} and the relation between gamma ray bursts and phase
transitions in stellar matter under rotation should also be extensively 
considered. The 
present work is the seed for this investigation since we consider only a few
possibilities for the EoS at $T=0$.

This paper is organized as follows: in Sec. II the formalisms of the
Hartle-Thorne approximation  are revisited and the results are presented. 
In Sec. III the results are discussed and the main conclusions are drawn.

\section{Formalism and Results}

As the first step we need to know the EoS of the system,
\[
\epsilon=\epsilon(p),\qquad n=n(p),
\]
where $p$ is the pressure, $\epsilon$ is the energy density, and $n$ is the 
number density of baryons. Once an 
adequate EoS is obtained, it can be used to provide the stellar properties.
{\bf If we opt for a non-rotating configuration, the
  Tolman-Oppenheimer-Volkoff (TOV) equations \cite{tov}, which are derived 
from Einstein's equations in the Schwarzschild metric for a static, 
spherical star are used to compute the stellar properties, as the radius, 
gravitational and baryonic masses and central energy densities.
If we opt for its rotating counterpart, the same EoS is used, but the 
Hartle-Thorne formalism is used instead of the TOV approximation.}

\subsection{Hartle-Thorne formalism}
The simplest nontrivial form that Einstein's equations take is the form for spherically symmetric stars. In this case, the line element has the form,
\begin{equation}
ds^2=e^{\nu(r)}dt^2+e^{\lambda(r)}dr^2+r^2(d\theta^2+\sin^2\theta\d\phi^2).
\end{equation}
However for rotating stars the spherical symmetry is broken, but the star maintains axial symmetry. The expression for the line element for axially symmetric spacetime is given by
\begin{eqnarray}
ds^2&=&e^{\nu(r,\theta)}dt^2-e^{\lambda(r,\theta)}dr^2-e^{\mu(r,\theta)}[r^2d\theta^2\nonumber\\
&+&r^2\sin^2\theta(d\phi-\omega(r,\theta)dt^2)^2].
\end{eqnarray}

The perfect fluid has a stress-energy tensor
\begin{eqnarray}
T^{ab}=(\epsilon+p)u^au^b+pg^{ab},
\end{eqnarray}
where $u^a$ is the 4-velocity of the fluid. 
By using (2) , (3) and the Ricci tensor we obtain a system of partial
differential equations in which the solution is numerically complicated. In
the Hartle-Thorne method, rotation is treated as a pertubation on the
non-rotating configuration of the star and it gives spherical and quadrupole
deformations. Within this method the expansion of the metric has the form
\begin{eqnarray}
e^{\lambda(r,\theta)}&=&e^{\lambda}\left[1+2\frac{m_0+m_2P_2(\cos(\theta))}{}\right],\nonumber\\
e^{\nu(r,\theta)}&=&e^{\nu}\left[1+2(h_0+h_2P_2(\cos(\theta)))\right],\\
e^{\mu(r,\theta)}&=&e^{\mu}\left[1+2(v_2-h_2)P_2(\cos(\theta))\right],\nonumber
\end{eqnarray}
where $P_2$ is the Legendre polynomial of second order; and $h_0$, $h_2$, $m_0$, $m_2$ and $v_2$ are all functions of $r$.

A non-rotating configuration is computed by integrating the
Tolman-Oppenheimer-Volkoff equations \cite{tov} of hydrostatic equilibrium for the pressure, $p(r)$, and the mass interior to a given radius, $m(r)$:
\begin{eqnarray}
\frac{dp}{dr}&=&-(\epsilon+p)\frac{m+4\pi r^3p}{r(r-2m)},\nonumber\\
\frac{dm}{dr}&=&4\pi r^2\epsilon,\\
\frac{d\nu}{dr}&=&-2(\epsilon+p)^{-1}\frac{dp}{dr},\nonumber
\end{eqnarray}
where the boundary conditions are $m=0$ at $r=0$, the radius surface $R$ is 
defined so that $\nu=0$ when $r\to \infty$. Here, $\nu$ is the metric function. In this work we use natural units where $G=c=1$.\\

 The quantity $\overline{\omega}=\Omega-\omega$ is the angular velocity of the
 fluid relative to the local inertial frame. In the classical mechanics the
 magnitude of the centrifugal force is determined by the angular velocity
 $\Omega$. However, in the general relativity the magnitude of the centrifugal force is determined by $\overline{\omega}$:
\begin{equation}
\frac{1}{r^4}\frac{d}{dr}\left(r^4j\frac{d\overline{\omega}}{dr}\right)+\frac{4}{r}\frac{dj}{dr}\overline{\omega}=0,
\end{equation}
where
\begin{equation}
j(r)=e^{\nu/2}[1-2m/r]^{1/2}.
\end{equation}
The boundary conditions $\overline{\omega}=\overline{\omega}_c$ and $d\overline{\omega}_c/dr=0$ are imposed. When $r=R$ we can determine the angular velocity, $\Omega$, and angular momentum, $J$, corresponding to $\overline{\omega}_c$:
\begin{equation}
J=\frac{1}{6}R^4\left(\frac{d\overline{\omega}}{dr}\right)_{r=R},\qquad
\Omega=\overline{\omega}(R)+\frac{2J}{R^3}.
\end{equation}
To obtain a different value of $\Omega$ we need to rescale the function $\overline{\omega}(r)$:
\begin{equation}
\overline{\omega}(r)_{new}=\overline{\omega}(r)_{old}(\Omega_{new}/\Omega_{old}).
\end{equation}
The metric function $h_0$ obeys the following algebric relations
\begin{eqnarray}
h_0&=&-\frac{m_0+J^2/r^3}{r-2m}+\frac{J^2}{r^3(r-2m)},\qquad\textrm{outside the star},\nonumber\\
h_0&=&-p_0+\frac{1}{3}r^2e^{-\nu}\overline{\omega}^2+h_{0c},\qquad\textrm{inside the star},
\end{eqnarray}
where $h_{0c}$ is a constant determined by imposing $h_0$ to be continuous at $r=R$. The mass pertubation factor $m_0$ and the pressure pertubation factor $p_0$ are calculated by integrating:
\begin{widetext}
\begin{eqnarray}
\frac{dm_0}{dr}&=&4\pi r^2\frac{d\epsilon}{dp}(\epsilon+p)p_0+\frac{1}{12}j^2r^4\left(\frac{d\overline{\omega}}{dr}\right)^2-\frac{1}{3}r^3\frac{dj^2}{dr}\overline{\omega}^2,\nonumber\\
\frac{dp_0}{dr}&=&-\frac{m_0(1+8\pi r^2p)}{(r-2m)^2}-\frac{4\pi(\epsilon+p)r^2}{(r-2m)}p_0+\frac{1}{12}\frac{r^4j^2}{(r-2m)}\left(\frac{d\overline{\omega}}{dr}\right)^2+\frac{1}{3}\frac{d}{dr}\left(\frac{r^3j^2\overline{\omega}^2}{r-2m}\right),
\end{eqnarray}
\end{widetext}
where the boundary conditions are that $m_0(0)=p_0(0)=0$.
The mass correction at first order is given by
\begin{eqnarray}
\delta m=m_0+J^2/R^3.\nonumber
\end{eqnarray}
Thus the total mass of a rotating neutron star with corrections up to the 
first order is given by
\begin{eqnarray}
m(R)+\delta m=m(R)+m_0(R)+J^2/R^3,\nonumber
\end{eqnarray}
where $R$ is the radius of the star surface.

According to \cite{ht97}, the binding energy of a relativistic star 
in a non-rotating configuration is the 
difference between its baryon mass and its total mass-energy
\begin{equation}
E_B=A-M,\nonumber
\end{equation}
where $A$ is the total baryon mass
\begin{equation}
A=m_n \int^R_0n(r)\left(1-\frac{2M}{r}\right)^{-1/2}4\pi r^2dr,
\end{equation}
and $m_n$ is the nucleon rest mass. To calculate the binding energy for the
non-rotating configuration we need to solve the  Tolman-Oppenheimer-Volkoff
equation. The first order correction in the binding energy for rotating 
configuration is given by
\begin{widetext}
\begin{eqnarray}
\delta E_B &=& -\frac{J^2}{R^3}+\int_0^R4\pi r^2B(r)dr,\nonumber\\
B(r)&=&(\epsilon+p)p_0\left[\frac{d\epsilon}{dp}\left(\left(1-\frac{2M}{r}\right)^{-1/2}-1\right)-\frac{d\epsilon_i}{dp}\left(1-\frac{2M}{r}\right)^{-1/2}\right]+(\epsilon-\epsilon_i)\left(1-\frac{2M}{r}\right)^{-3/2}\left[\frac{m_0}{r}+\frac{1}{3}j^2r^2\overline{\omega}^2\right]\nonumber\\
&-&\frac{1}{4\pi r^2}\left[\frac{1}{12}j^2r^4\left(\frac{d\overline{\omega}}{dr}\right)^2-\frac{1}{3}\frac{dj^2}{dr}r^3\overline{\omega}^2\right],
\end{eqnarray}
\end{widetext}
where $\epsilon_i$ is the internal energy density,
\begin{equation}
\epsilon_i=\epsilon- m_n n.\nonumber
\end{equation}
Thus the baryon mass of rotating stars is given by
\begin{equation}
M_B=A+\delta E_B+ \delta m.
\end{equation}

The metric functions $h_2$ and $v_2$ are given by
\begin{widetext}
\begin{eqnarray}
\frac{dv_2}{dr}&=&\left(\frac{1}{r}+\frac{1}{2}\frac{d\nu}{dr}\right)\left[-\frac{1}{3}r^3\frac{dj^2}{dr}\overline{\omega}^2+\frac{1}{6}j^2r^4\left(\frac{d\overline{\omega}}{dr}\right)^2\right]-\frac{d\nu}{dr}h_2,\nonumber\\
\frac{dh_2}{dr}&=&\left[-\frac{d\nu}{dr}+\frac{r}{r-2m}\left(\frac{d\nu}{dr}\right)^{-1}\left[8\pi(\epsilon+p)-\frac{4m}{r^3}\right]\right]h_2-\frac{4v_2}{r(r-2m)}\left(\frac{d\nu}{dr}\right)^{-1}-\frac{1}{3}\left[\frac{1}{2}\frac{d\nu}{dr}+\frac{1}{r-2M}\left(\frac{d\nu}{dr}\right)^{-1}\right]r^3j^2\left(\frac{d\overline{\omega}}{dr}\right)^2\nonumber\\
&+&\frac{1}{6}\left[\frac{1}{2}\frac{d\nu}{dr}r-\frac{1}{r-2m}\left(\frac{d\nu}{dr}\right)^{-1}\right]r^2\frac{dj^2}{dr}\overline{\omega}^2,
\end{eqnarray}
\end{widetext}
where the boundary conditions are $h_2(0)=v_2(0)=0$.\\
The non-radial mass and pressure perturbation factors, $m_2$ and $p_2$, are determined from the algebric relations
\begin{eqnarray}
m_2&=&(r-2m)\left[-h_2-\frac{1}{3}r^3\left(\frac{dj^2}{dr}\right)\overline{\omega}^2+\frac{1}{6}r^4j^2\left(\frac{d\overline{\omega}}{dr}\right)^2\right],\nonumber\\
p_2&=&-h_2-\frac{1}{3}r^2e^{-\nu}\overline{\omega}^2.
\end{eqnarray}
To calculate the eccentricity $e$ we use
\begin{equation}
e=\left(1-\frac{R_p^2}{R_e^2}\right)^{1/2},
\end{equation}
where $R_p$ and $R_e$ are the polar and equatorial radii. To obtain $R_p$ and $R_e$ we use the relations
\begin{eqnarray}
S(\theta)&=&r+\xi_0(r)+\xi_2(r)P_2(cos(\theta)),\nonumber\\
\xi_0&=&-p_0(\epsilon+p)/(dp/dr),\\
\xi_2&=&-p_2(\epsilon+p)/(dp/dr).\nonumber
\end{eqnarray}
where $S(\theta)$ is a surface of constant density.

In table \ref{tab:tab1} results for slowly rotating stars are presented. The 
$M_{max}$ and $R$ are respectively the maximum mass and the radius of an 
analogous non-rotating configuration. The $M_{max}^1$ is the maximum mass 
corrected up to first order for a star with angular velocity $\Omega$. 
$R$ is the radius of the non-rotating star, $R_e$ is the equatorial radius, 
$R_p$ is the polar radius, $\epsilon_c$ is the assumed central energy density,
$I$ is the calculated moment of inertia and $e$ is the eccentricity.
The EoS for hadronic and hybrid stars were taken from \cite{trapping} and the
EoS for quark stars were taken from \cite{quark}. As these EoS have
already been extensively discussed in the literature, we refrain from further
explanations here. The only point worth mentioning refers to the 
parametrizations used. For hadronic and hybrid stars we have chosen to work 
with a parametrization that describes the properties of saturating nuclear 
matter proposed in \cite{Glen00} since other common parameter sets for the 
non-linear Walecka model namely, TM1 \cite{tm1} and NL3 \cite{nl3} proved 
to be inadequate because, due to the inclusion of hyperons,  the nucleon mass 
becomes negative at relatively low densities \cite{magno,mp1}. The chosen 
parameters are $g_s^2/m_s^2=11.79$ fm$^2$, $g_v^2/m_v^2=7.148$ fm$^2$,
$g_{\rho}^2/m_{\rho}^2=4.41$ fm$^2$, $\kappa/M=0.005896$ and 
$\lambda=-0.0006426$, for which the binding energy is -16.3 MeV at the 
saturation density $\rho_0=0.153$ fm$^{-1}$, the symmetry coefficient is 
32.5 MeV, the compression modulus is 300 MeV and the effective mass is $0.7 M$.
For the meson-hyperon coupling constants we have chosen them 
constrained by the binding of the $\Lambda$ hyperon in nuclear matter, 
hypernuclear levels and neutron star masses
($x_\sigma=0.7$ and $x_\omega=x_\rho=0.783$) and have assumed that the 
couplings to the $\Sigma$ and $\Xi$ are equal to those of the $\Lambda$
hyperon \cite{gm91,Glen00}. 
For the construction of the EoS for hybrid stars, the hadronic phase was
obtained with the non-linear Walecka model and the parameters above and the
quark phase with the MIT bag model with $Bag=(180{\,\ MeV})^4$.
For the quark stars within the MIT model,
we have used $m_u=m_d=5.5\rm\,MeV$, $m_s=150.0\rm\,MeV$ and $Bag=(180{\,\rm 
MeV})^4$. For the purpose of the present work, the value of the bag parameter
does not play an important role. Concerning quark stars within the NJL
model, the set of parameters were chosen in order to fit the values in vacuum 
for the pion mass, the pion decay constant,  the kaon mass and the 
quark condensates \cite{Ruivo99,kun89}:  $\Lambda=631.4\rm\,MeV$, $ 
g_S\,\Lambda^2=1.824$,  $g_D\,\Lambda^5=-9.4$,
$m_u=m_d=5.6\rm\,MeV$ and $m_s=135.6\rm\,MeV$  which were fitted to 
the following properties: $m_\pi=139\rm\,MeV$, $f_\pi=93.0\rm\,MeV$, 
$m_K=495.7\rm\,MeV$, $f_K=98.9\rm\,MeV$, $\langle\bar u 
u\rangle=\langle\bar d d\rangle=-(246.7\,\mbox{ MeV})^3$ and 
$\langle\bar s s\rangle=-(266.9\,\mbox{ MeV})^3$.

One can easily see from table \ref{tab:tab1} that for all models considered, 
the rotating stars bears a slightly larger mass than its non-rotating
counterpart. Eccentricities are rather uniform for all models (0.43-0.49). 
Quark stars are bound by the strong force and hence they can rotate faster
than hadronic or hybrid stars that are bound by the gravitational force. 
According to \cite{haensel2}, the value of the Kepler frequency can be
obtained from the values of the mass and the radius of the corresponding
non-rotating star and its empirical relation reads
\begin{equation}
\Omega=(0.63-0.65) (M/R^3)^{1/2}.
\end{equation}
In table \ref{tab:tab1} the Kepler frequency for each EoS is specified. While
hadronic and hybrid stars rotate with similar frequencies, the values for
quark stars are indeed much higher.
  
\begin{table*}[h]
\begin{center}
\caption{\label{tab:tab1}Rotating compact stars properties with Kepler angular velocity $\Omega=0.65(M/R^3)^{1/2}$.}
\begin{ruledtabular}
\begin{tabular}{ccccccccccc}
\hline
Type & Entropy & $M_{\max}$ & $ M_{\max}^{1} $ & $R$ & $R_e$ & $R_p$ & 
$\epsilon_c$ & $I$ & e & $\Omega$\\
&& $(M_{\odot})$ & $(M_{\odot})$ & $(km)$ & $(km)$  & $(km)$  & $(g/cm^3)$ & $(gcm^2)$ &  & $ Hz $\\
\hline
Hadronic       & 0 & 2.04 & 2.07 & 11.73 & 12.06 & 10.86 & $1.98\times 10^{15}$ & $2.39\times 10^{45}$ & 0.43 & 1661.37 \\
Hadronic       & 1 & 1.96 & 2.00 & 11.02 & 11.33 & 10.21 & $2.23\times 10^{15}$ & $2.11\times 10^{45}$ &0.43 & 1791.66  \\
Hadronic       & 2 & 1.93 & 1.96 & 10.91 & 11.22 & 10.11 & $2.24\times 10^{15}$ & $2.03\times 10^{45}$ &0.43 & 1801.74  \\
\hline
Hybrid         & 0 & 1.63 & 1.65 & 12.33 & 12.78 & 11.19 & $1.57\times 10^{15}$ & $1.90\times 10^{45}$ & 0.48 & 1381.09 \\
Hybrid         & 1 & 1.50 & 1.53 & 11.32 & 11.73 & 10.28 & $1.75\times 10^{15}$ & $1.56\times 10^{45}$ & 0.48 & 1503.07 \\
Hybrid         & 2 & 1.50 & 1.52 & 11.76 & 12.20 & 10.65 & $1.58\times 10^{15}$ & $1.65\times 10^{45}$ & 0.49 & 1418.80 \\
\hline
Quarkonic(MIT) & 0 & 1.22 & 1.25 &  6.76 & 6.97 & 6.23 & $5.14\times 10^{15}$ & $0.53\times 10^{45}$ & 0.45 & 2938.89 \\
Quarkonic(MIT) & 1 & 1.22 & 1.25 &  6.76 & 6.97 & 6.23 & $5.17\times 10^{15}$ & $0.53\times 10^{45}$ & 0.45 & 2941.47 \\
Quarkonic(MIT) & 2 & 1.23 & 1.26 &  6.79 & 6.99 & 6.25 & $5.08\times 10^{15}$ & $0.54\times 10^{45}$ & 0.45 & 2929.40 \\
\hline
Quarkonic(NJL) & 0 & 1.20 & 1.23 & 7.87 & 8.12 & 7.24 & $3.45\times 10^{15}$ & $0.69\times 10^{45}$ & 0.45 & 2316.71 \\
Quarkonic(NJL) & 1 & 1.17 & 1.18 &  7.71 & 7.81 & 7.45 & $3.68\times 10^{15}$ & $0.64\times 10^{45}$ & 0.45 & 2363.51 \\
Quarkonic(NJL) & 2 & 1.10 & 1.13 &  7.18 & 7.39 & 6.62 & $4.58\times 10^{15}$ & $0.51\times 10^{45}$ &0.44 & 2552.36 \\
\hline
\end{tabular}
\end{ruledtabular}
\end{center}
\end{table*}

\section{Results and Conclusions}

Let's now go back to our results in order to compare them with what is found
in the literature and draw the conclusions. 

We first analyzed the results of the rotating pulsars.
The values shown in table \ref{tab:tab1} for zero entropy and hadronic stars 
show the same behavior
observed in \cite{antigo2}, i.e., there is a small increase in the maximum
mass of the rotating star as compared with the non-rotating configuration.
Notice, however, that in \cite{antigo2}, the usual
NL3 and TM1 parametrizations could be used because only protons and neutrons 
were considered and the central density is somewhat lower than in our case.

In \cite{antigo1} the
models named O,P,Q and R given in tables \ref{tab:tab1} can again be compared 
with our result for the hadronic star at $T=0$ and they are indeed very
similar. The central density and the moment of inertial are of the same order
and the maximum mass is very similar. As far as we know there is no result for
rotating stars with entropy different from zero (finite temperature) in the
literature, but we can examine its effect on the stellar properties from
table \ref{tab:tab1}. As entropy increases the maximum masses decrease for all
kinds of pulsars, except for the not so precise MIT bag model. 
Although the radii of a non-rotating star degenerally
decreases with temperature (except again for the very simple and unrealistic 
MIT model), the eccentricity of a rotating star remains practically unchanged.
As expected, the moments of inertia of quark stars are much lower than of
hadronic and hybrid stars with consequent larger rotation frequencies.

The energy released in the conversion mechanism of the metastable (MS) star 
into a stable star (SS) is given by
\begin{equation}
\Delta E = [(M_G(MS)-M_G(SS))/M_{\odot}]\times 17.88\times 10^{53}erg.\nonumber
\end{equation} 
where $M_G(MS)$ is the gravitational mass of the metastable star and $M_G(SS)$ 
is the gravitational mass of the stable star. For $\Delta E$ to be positive, 
the gravitational mass of the metastable star, at a fixed baryonic mass, has 
to be larger than the gravitational mass of the stable star.
We have calculated the  released energies in the conversion mechanism for 
the hadronic to the hybrid, hadronic to the quark and hybrid to the
quark stars with $S=0$ in the rotating configuration at fixed baryon mass. 
The released energy is always negative, except 
in the conversion mechanism of the hadronic (MS) to the hybrid star (SS) 
at a fixed baryonic mass of $1.56M_{\odot}$, which yields  
$\Delta E = 1.14\times 10^{51}erg$, allowing for the required energy measured
in a short gamma ray burst.

In the present work we have investigated one possible variation on the usual
static neutral pulsars: the inclusion of rotation. 
We have observed that the behaviors shown in previous works with much
simpler EoS were also observed here. The influence of the temperature was also
investigated in both cases. We are now in a position to calculate the energy
released from the conversion of a metastable star (hadronic or hybrid) to a 
stable star (hybrid or quark) under the influence of slow rotation.
For the simple calculations done so far with stars in a
rotating configuration at $S=0$, only a hadronic star could convert into a
hybrid star releasing an energy compatible with a short gamma ray burst.
A more detailed and complete investigation, with a smaller bag parameter 
in the MIT bag model is under way.

\section*{Acknowledgments}
This work was partially supported by CNPq (Brazil). M.D.A. would like to thank
CNPq for a master's degree scholarship.

\end{document}